\documentclass[10pt,aps,prd,nofootinbib,superscriptaddress,twocolumn]{revtex4}
\usepackage[utf8]{inputenc}
\DeclareUnicodeCharacter{200B}{{\hskip 0pt}}
\usepackage{color}
\usepackage{graphicx}
\usepackage{amsmath}
\usepackage{amssymb}
\usepackage{bm}
\usepackage{acronym}
\usepackage{ifthen}
\usepackage{blindtext}
\usepackage[normalem]{ulem}
\usepackage{hyperref}
\usepackage{etoolbox}
\usepackage{fancyhdr}
\usepackage{xspace}
\usepackage{textcomp}
\usepackage{multirow}
\usepackage[caption=false]{subfig}
\usepackage{lineno}
\usepackage{tabularx}
\hypersetup{
    colorlinks=true,
    linkcolor=blue,
    filecolor=magenta,      
    urlcolor=black,
		citecolor=red,
		}
\begin{document}

\title{On the Gravitational Angular Momentum of Axial Perturbations of a Regular Black Hole}

\author{S. C. Ulhoa}
\email{sc.ulhoa@gmail.com}
\affiliation{Instituto de F\'isica, Universidade de Bras\'ilia, 70910-900, Bras\'ilia, DF, Brazil}
\affiliation{Canadian Quantum Research Center, 204-3002 32 Ave, Vernon, BC V1T 2L7, Canada
}

\author{F. L. Carneiro}\email{fernandolessa45@gmail.com}
\affiliation{Universidade Federal do Norte do Tocantins, 77824-838, Aragua\'ina, TO, Brazil}

\author{B. C. C. Carneiro}
\email{bcccarneiro@gmail.com
}
\affiliation{Instituto Federal do Tocantins, 77760-000, Colinas do Tocantins, TO, Brazil}

\begin{abstract}
This Letter deals with the gravitational angular momentum carried by axial
(odd-parity) perturbations of the Bardeen regular black hole within the
teleparallel equivalent of general relativity (TEGR). Using the Hamiltonian
definition of conserved quantities in TEGR, we derive a closed expression for
the perturbative angular momentum $\delta J$ in terms of the axial perturbation
function $h_0(r,t)$. The result exhibits a sharp multipolar selection rule:
$\delta J$ vanishes for odd values of the multipole index $\ell$, while even-$\ell$
modes yield a nonzero contribution. The radial and temporal behavior of $\delta J$ is
illustrated using the known axial quasinormal modes of the Bardeen spacetime.
\end{abstract}


\maketitle

\date{\today}

\section{Introduction}

Observations of compact astrophysical objects have reached an unprecedented level
of detail \cite{EHT2019,EHT2022}, yet they do not provide direct information about
the internal structure of the strong-field region. In particular, current observational data are consistent with a variety of black-hole models and do not allow one to discriminate between singular solutions of General Relativity and nonsingular alternatives. This intrinsic observational limitation motivates the investigation of regular black-hole spacetimes as viable candidates for describing extremely compact objects in nature.

Among nonsingular alternatives, regular black-hole solutions have attracted
considerable attention since their original proposal by Bardeen~\cite{Bardeen1968},
as they evade curvature singularities while preserving an event horizon.
For these geometries to represent physically viable compact objects, their
stability under \emph{gravitational} perturbations is a crucial requirement.
While earlier studies of regular black holes focused mainly on perturbations of
test fields or of the matter sector sourcing the geometry, the stability analysis
under genuine gravitational perturbations was first carried out for the Bardeen
spacetime in Ref.~\cite{Ulhoa2014}. In that work, an axial (odd-parity)
Regge--Wheeler--type master equation was derived and the corresponding quasinormal
spectrum was obtained, establishing the dynamical consistency of this regular
black-hole solution.

Gravitational perturbations encode the dynamical response of a black hole to
external disturbances and thus carry physically meaningful information.
In particular, conserved quantities such as energy and angular momentum provide
a direct characterization of how gravitational degrees of freedom propagate and
interact. Within the Teleparallel Equivalent of General Relativity (TEGR), these
quantities arise as well-defined surface integrals associated with spacetime
symmetries, yielding unambiguous expressions for gravitational conserved charges~ \cite{Maluf2013}.

In the Hamiltonian formulation of TEGR, the gravitational energy--momentum vector
$P^{a}$ and the angular momentum tensor $L^{ab}$ emerge as generators of spacetime
translations and global Lorentz transformations, respectively. These quantities
satisfy the Poincar\'e algebra and therefore provide a physically consistent
definition of conserved gravitational charges. In the present work, $\delta J$ is interpreted within the Hamiltonian framework of TEGR as a gravitational quantity associated with the axial perturbative sector of the Bardeen spacetime. In this sense, it is not introduced here from a standard flux-based or quasi-local construction. Although a comparison with more conventional asymptotic or radiative notions of angular momentum would be of interest, such an analysis would considerably broaden the scope of the present Letter and is therefore left for future work. In this Letter, we employ this
framework to investigate the gravitational angular momentum carried by axial
perturbations of the Bardeen regular black hole, taking the known quasinormal
modes of the axial sector as input. Our main result shows that a nonvanishing
gravitational angular momentum arises only for even multipolar index $\ell$,
while odd modes do not contribute.

\section{Gravitational Angular Momentum of Axial Perturbations in the Bardeen Spacetime}

We consider the Bardeen regular black hole as the background spacetime, described
by the spherically symmetric line element
\begin{equation}
ds^2
=
-f(r)\,dt^2
+
\frac{1}{f(r)}\,dr^2
+
r^2 d\theta^2
+
r^2\sin^2\theta\,d\phi^2 ,
\end{equation}
where
\begin{equation}
f(r)
=
1-\frac{2Mr^2}{(r^2+\alpha^2)^{3/2}} \,,
\end{equation}
where $M$ is the black hole mass and $\alpha$ characterizes the regularization scale of the Bardeen geometry. Following Ref.~[4], we consider axial (odd-parity) gravitational perturbations of the Bardeen metric, so that the functions $h_0(r,t)$ and $h_1(r,t)$ introduced below represent the corresponding axial metric perturbations. Their dynamics is governed by the linearized axial perturbation equations derived in Ref.~[4], which are taken here as the starting point for the TEGR angular-momentum calculation. The perturbations are introduced through the metric components
\begin{equation}
\delta g_{0\phi}
=
\epsilon\,h_0(r,t)\,P_\ell(\cos\theta),
\qquad
\delta g_{r\phi}
=
\epsilon\,h_1(r,t)\,P_\ell(\cos\theta),
\end{equation}
where $P_\ell$ denotes the Legendre polynomial of degree $\ell$ and $\epsilon$
controls the perturbative expansion.

In the Hamiltonian formulation of the Teleparallel Equivalent of General
Relativity~\cite{Maluf2013}, the gravitational angular momentum is encoded in the Lorentz
generator $L^{ab}$, which is defined as the volume integral of the gravitational
angular momentum density $M^{ab}$,
\begin{equation}
L^{ab}=-\int_V d^3x\,M^{ab}.
\end{equation}
By rearranging the antisymmetric structure of the Hamiltonian density, the angular momentum density can be written in the compact form
\begin{equation}
M^{ab}
=
-k\,\varepsilon^{ab}{}_{cd}\,\varepsilon^{ijk}\,
\partial_i\!\left(
e^{c}{}_{j}\,e^{d}{}_{k}
\right),
\qquad
k=\frac{1}{16\pi},
\end{equation}
where $e^{a}{}_{\mu}$ denotes the tetrad field. Here $k$ is the gravitational coupling constant in natural units. The gravitational angular momentum vector is obtained from the rotational sector
of $L^{ab}$. Restricting to spatial indices, its three components are defined as
\begin{equation}
J^{(i)}=\frac{1}{2}\,\varepsilon^{(i)}{}_{(j)(k)}\,L^{(j)(k)},
\qquad (i,j,k=1,2,3).
\end{equation}
Using the contraction of Levi-Civita symbols restricted to the spatial sector,
this definition leads to the explicit expression
\begin{equation}
J^{(i)}
=
2k
\int_V d^3x\;
\epsilon^{ijk}\,
\partial_i\!\left(
e^{(0)}{}_j\,e^{(i)}{}_k
\right).
\end{equation}
This form shows that the angular momentum is entirely determined by a surface
term constructed from the tetrad components, which can be directly evaluated
once a specific tetrad adapted to the background geometry is chosen.

In order to evaluate the gravitational angular momentum explicitly, we adopt a tetrad
field adapted to static observers in the Bardeen spacetime. We follow the
construction introduced in Ref.~\cite{MalufUlhoa2021}, which has proven
convenient for the analysis of gravitational perturbations in the teleparallel
framework. The tetrad is given by
\begin{widetext}
\begin{equation}
e_{a\mu}
=
\begin{pmatrix}
 -A & 0 & 0 & -D \\
 0 & B \sin\theta \cos\phi & r \cos\theta \cos\phi &
     - E r \sin\theta \sin\phi + F \sin\theta \cos\phi \\
 0 & B \sin\theta \sin\phi & r \cos\theta \sin\phi &
     E r \sin\theta \cos\phi + F \sin\theta \sin\phi \\
 0 & B \cos\theta & - r \sin\theta & F \cos\theta
\end{pmatrix},
\end{equation}
where the background functions are defined as
\begin{equation*}
A=\sqrt{f(r)},
\qquad
B=\frac{1}{\sqrt{f(r)}},
\qquad
E^2
=
1+\frac{g_{11}g_{03}^2+g_{00}g_{13}^2}{r^2\sin^2\theta},
\qquad
F=g_{13}\sqrt{g_{11}}.
\end{equation*}
\end{widetext}
For axial perturbations, the only tetrad component receiving a linear
contribution is
\begin{equation}
D=-\frac{g_{03}}{\sqrt{-g_{00}}}
=
-\epsilon\,\frac{h_0(r,t)}{A(r)}\,P_\ell(\cos\theta),
\end{equation}
while all remaining functions coincide with their background values. As a
result, the gravitational angular momentum is linear in the perturbative
parameter $\epsilon$, and its evaluation reduces to the explicit computation
of the corresponding surface term. The tetrad in Eq.~(8) is fixed by the choice of static observers and by the form of the line element, with the metric functions entering only through its components. In the present framework, the tetrad is interpreted as a frame adapted to a congruence of observers. In particular, the observer four-velocity is identified with the timelike component of the inverse tetrad, namely $U^{\mu}=e_{(0)}{}^{\mu}=(1/A,0,0,0)$, which shows that the chosen frame corresponds to static observers in the Bardeen spacetime. Since the Bardeen background has the same static, spherically symmetric structure as the Schwarzschild geometry, differing only in the explicit form of the metric function, the construction adopted from Ref.~\cite{MalufUlhoa2021} is directly applicable here. Moreover, in the trivial limit in which the perturbation is turned off and the gravitational field is absent, the corresponding torsion vanishes, as expected. Since the TEGR gravitational angular momentum is defined with respect to the chosen reference frame, the result obtained here refers to the static observers associated with Eq.~(8). In particular, the even/odd multipolar selection rule established below is derived within this specific frame choice.

Due to the explicit $\phi$--dependence of the tetrad components, the integrands
associated with $J^{(1)}$ and $J^{(2)}$ are proportional to $\cos\phi$ and
$\sin\phi$, respectively. Upon integration over the azimuthal angle, these
contributions vanish identically, implying that the gravitational angular
momentum has no transverse components. Consequently, only the axial component of the gravitational angular momentum
can be nonvanishing. Substituting the explicit form of the tetrad into the
general expression for $J^{(i)}$, one finds that the remaining component reduces
to
\begin{align}
J^{(3)}
&= 2k \int dr\, d\theta\, d\phi \;
\bigg[
\partial_r \bigl( D r \sin\theta \bigr)
\nonumber \\
&\qquad
+ \partial_\theta \bigl( D B \cos\theta \bigr)
\bigg],
\end{align}
which provides the starting point for the evaluation of the angular momentum
carried by axial gravitational perturbations.

Since the background Bardeen spacetime is static and spherically symmetric, its
gravitational angular momentum vanishes identically. We therefore define the
perturbative angular momentum as the axial component of $J^{(3)}$ at first order
in the perturbation parameter $\epsilon$,
\begin{equation}
\delta J
\equiv
\frac{J^{(3)}-J^{(3)}_{\mathrm{bg}}}{\epsilon},
\end{equation}
where $J^{(3)}_{\mathrm{bg}}=0$. Substituting the explicit form of $D$ and keeping
only linear terms in $\epsilon$, the angular momentum reduces to
\begin{align}
\delta J
&= -4\pi k \int dr\, d\theta \;
\Biggl[
  \sin\theta \, P_\ell(\cos\theta) \,
  \partial_r\!\left( \frac{r h_0}{A} \right)
\nonumber\\[1.2ex]
&\qquad
+ \frac{B h_0}{A} \,
  \partial_\theta\!\Bigl( \cos\theta \, P_\ell(\cos\theta) \Bigr)
\Biggr],
\end{align}
which depends explicitly on the multipole index $\ell$ through the angular
structure of the axial perturbations. The remaining angular integrations can be carried out analytically using
standard properties of Legendre polynomials. In particular, one has
\begin{equation*}
\int_0^\pi
\sin\theta\,P_\ell(\cos\theta)\,d\theta
=
2\,\delta_{\ell 0},
\end{equation*}
and
\begin{equation*}
\int_0^\pi
\partial_\theta\!\big(\cos\theta\,P_\ell(\cos\theta)\big)\,d\theta
=
-\big(1+(-1)^\ell\big).
\end{equation*}
Substituting these results into the expression for $\delta J$, the angular
momentum reduces to the purely radial form
\begin{equation}
\delta J
=
-4\pi k
\int dr
\left[
2\,\delta_{\ell 0}\,
\partial_r\!\left(\frac{r h_0}{A}\right)
-
\big(1+(-1)^\ell\big)
\frac{B h_0}{A}
\right].
\end{equation}
This expression explicitly displays the multipolar selection rules governing
the gravitational angular momentum carried by axial perturbations. Since the angular momentum is defined as a volume integral over a spatial
hypersurface, the radial integration is performed with respect to the
coordinate $r$, rather than the tortoise coordinate. Denoting by $R$ the radius
of the integration volume $V$, the contribution $\ell=0$ becomes
\begin{equation}
\delta J_{\ell=0}
=
-\frac{1}{2}\left.
\frac{r\,h_0(r,t)}{\sqrt{f(r)}}
\right|_{0}^{R}
+
\frac{1}{2}\int_{0}^{R} dr\;\frac{h_0(r,t)}{f(r)} .
\end{equation}
For regular black holes, the lower limit $r=0$ is nonsingular and does not
introduce additional contributions. For $\ell\geq 1$, the term proportional to $\delta_{\ell 0}$ vanishes. As a consequence, axial perturbations with odd $\ell$ do not carry angular
momentum, whereas even--$\ell$ modes yield
\begin{equation}\label{deltaJ}
\delta J_{\ell>0}
=
\begin{cases}
\displaystyle
\frac{1}{2}\int_{0}^{R} dr\;\dfrac{h_0(r,t)}{f(r)}, & \ell \ \text{even},\\[10pt]
0, & \ell \ \text{odd}.
\end{cases}
\end{equation}
These expressions make explicit that the gravitational angular momentum carried
by axial perturbations is entirely determined by the background metric function
and by the parity of the multipole index $\ell$. The selection rule in Eq.~(\ref{deltaJ}) follows directly from the angular structure of the perturbative expression for $\delta J$, in particular from the parity properties of the Legendre polynomials that enter the angular integration. In this sense, the vanishing of odd-$\ell$ contributions is a consequence of the axial angular dependence of the perturbation within the TEGR angular-momentum definition adopted here, rather than an independent dynamical property of the Bardeen background. Equation~(\ref{deltaJ}) is obtained in the formal domain of the angular integration and therefore displays the full dependence on the index $l$. However, when discussing the quasinormal-mode spectrum, we follow Ref.~\cite{Ulhoa2014}, where the behavior of the low-$l$ modes indicates that the fundamental mode is more appropriately identified with $\{l,n\}=\{2,0\}$, in analogy with the Schwarzschild case.

The closed expression obtained in Eq.~(\ref{deltaJ}) allows a direct evaluation of the gravitational angular momentum carried by axial perturbations once the quasinormal-mode solutions for $h_0(r,t)$ are specified. In order to illustrate the radial behavior of $\delta J$ and its time dependence for selected modes, we evaluate the integral using the axial quasinormal-mode functions of the Bardeen spacetime previously obtained in Ref.~\cite{Ulhoa2014}, within the third-order WKB approximation. The corresponding profiles are shown in Figs.~1--4. These plots should be understood as illustrations of $\delta J$ for the chosen quasinormal-mode contributions, rather than as the result of an evolution from specified initial data. In all cases, the radial coordinate $R$ is chosen to start at $R=4$, ensuring that the behavior of $\delta J$ is evaluated strictly outside the event horizon.

\begin{figure}[t]
\centering
\includegraphics[width=\columnwidth]{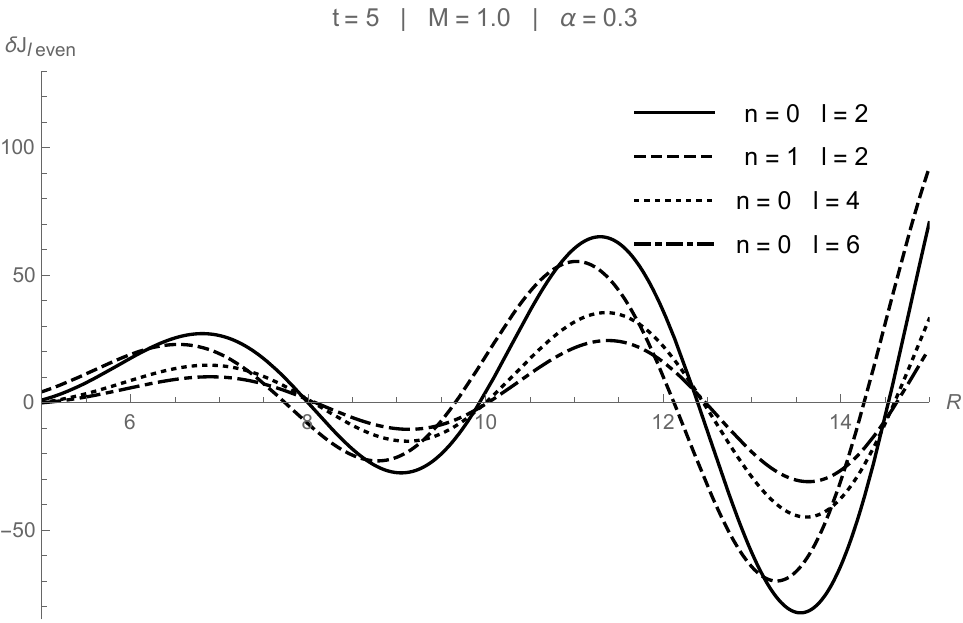}
\caption{Radial dependence of the perturbative gravitational angular momentum $\delta J$ for different even axial modes characterized by the multipolar index $\ell$ and overtone number $n$, evaluated at fixed time $t=5$, mass $M=1$, and regularization parameter $\alpha=0.3$.}
\label{fig:deltaJ_radius_modes}
\end{figure}

\begin{figure}[t]
\centering
\includegraphics[width=\columnwidth]{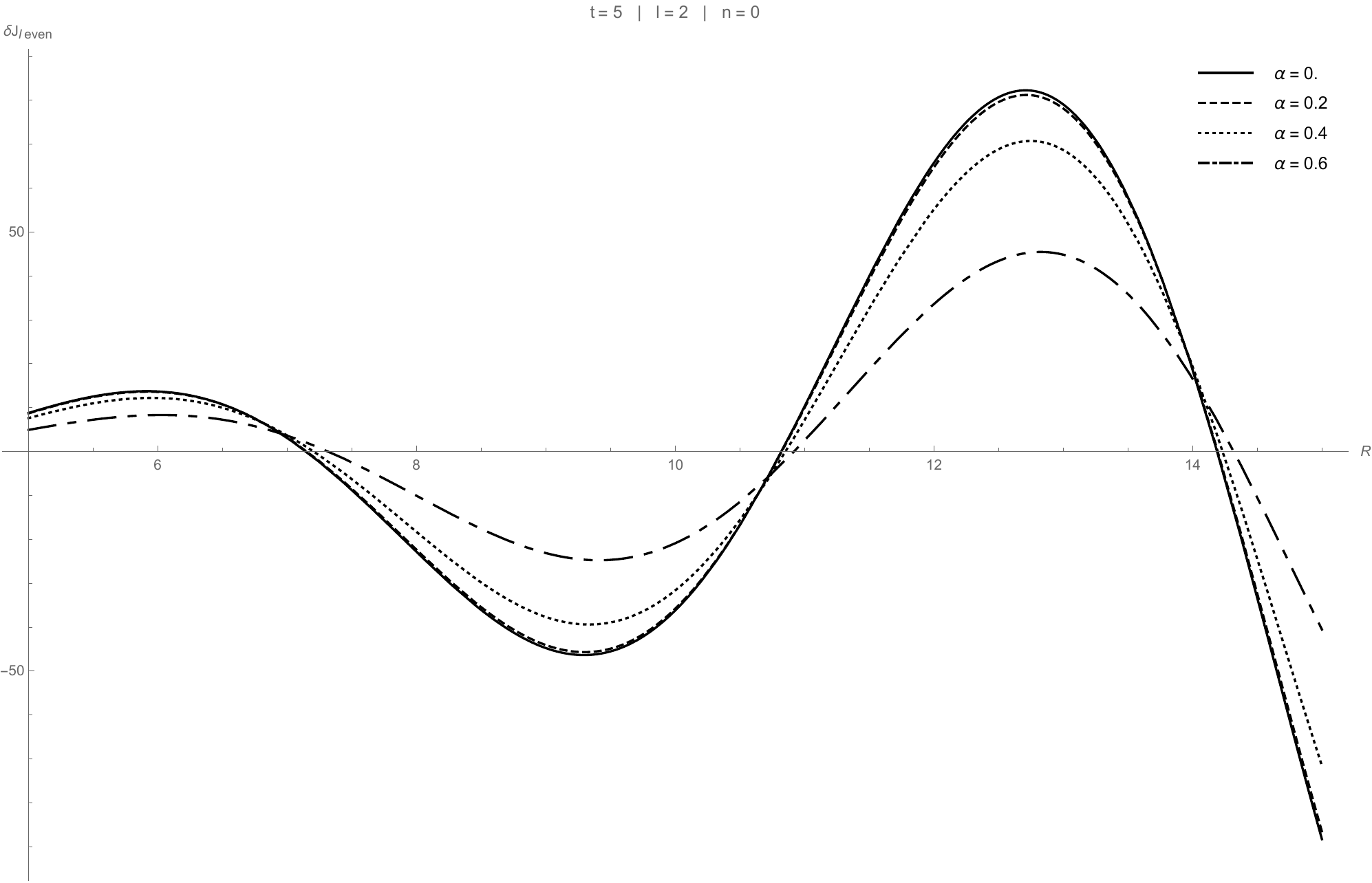}
\caption{Radial behavior of the perturbative gravitational angular momentum $\delta J$ for the fundamental axial mode $(\ell=2,n=0)$ evaluated at fixed time $t=5$ and mass $M=1$, for different values of the Bardeen regularization parameter $\alpha$.}
\label{fig:deltaJ_radius_alpha}
\end{figure}

\begin{figure}[t]
\centering
\includegraphics[width=\columnwidth]{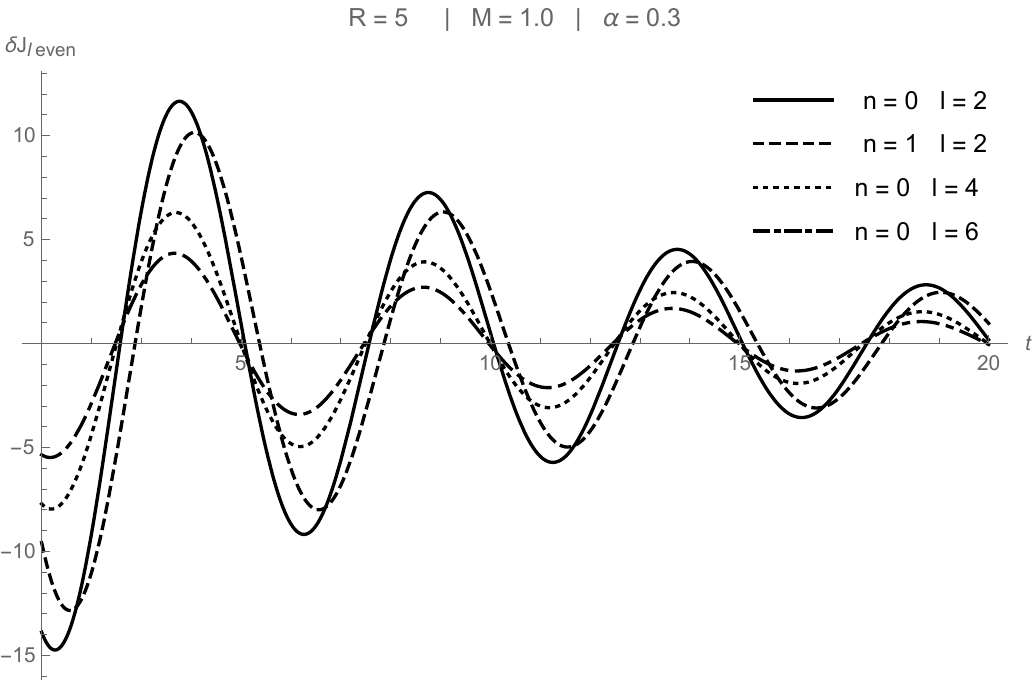}
\caption{Time evolution of the gravitational angular momentum $\delta J$ for different even axial quasinormal modes $(\ell,n)$, evaluated at fixed radius $R=5$, mass $M=1$, and $\alpha=0.3$.}
\label{fig:deltaJ_time_modes}
\end{figure}

\begin{figure}[t]
\centering
\includegraphics[width=\columnwidth]{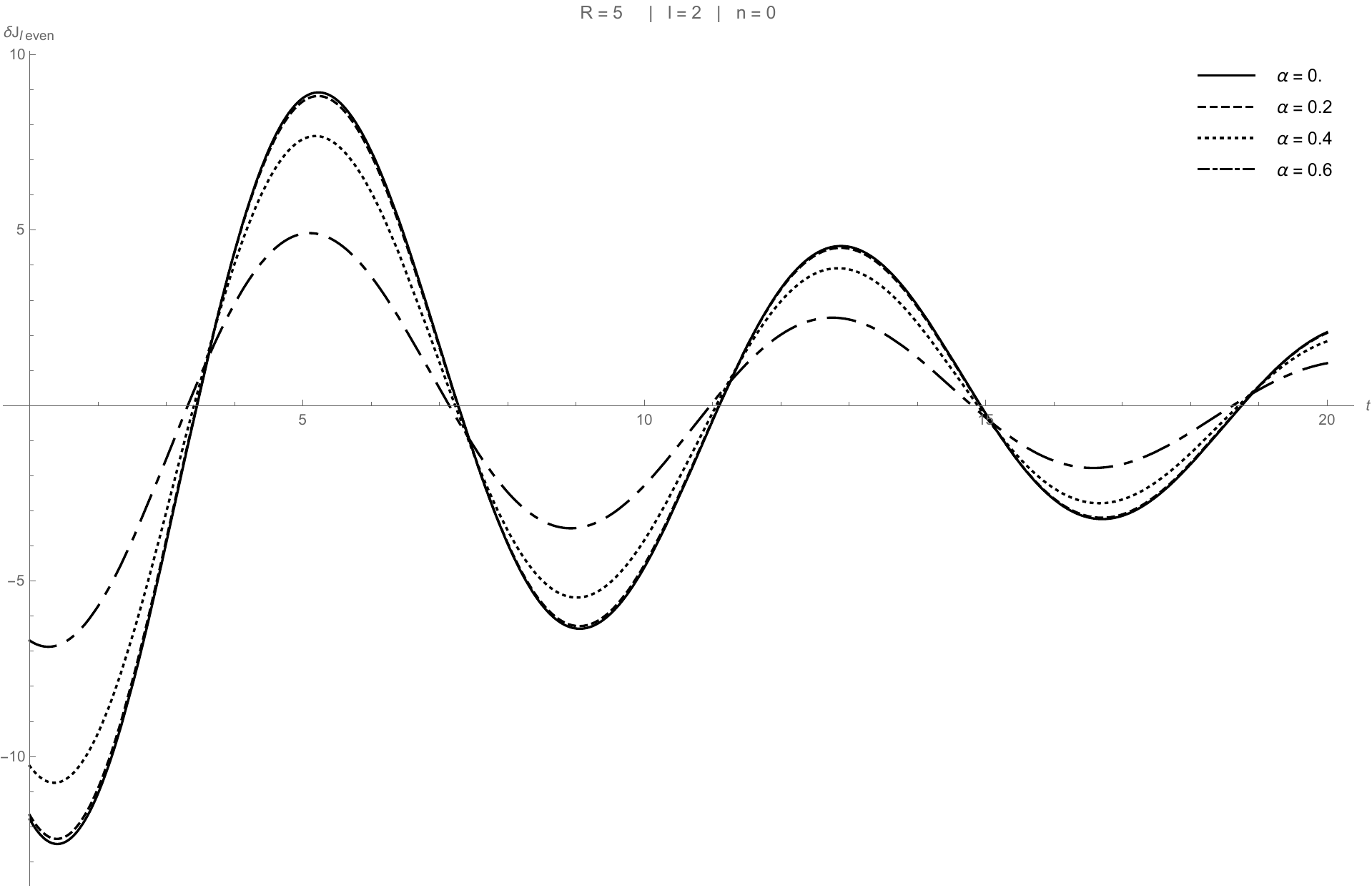}
\caption{Time dependence of the perturbative gravitational angular momentum $\delta J$ for the fundamental axial mode $(\ell=2,n=0)$ evaluated at fixed radius $R=5$ and mass $M=1$, for different values of the regularization parameter $\alpha$.}
\label{fig:deltaJ_time_alpha}
\end{figure}

In the plots shown here, no independent physical normalization condition was imposed on the perturbation amplitude. Accordingly, the absolute magnitude of $\delta J$ should not be overinterpreted in the present context. The purpose of Figs.~1--4 is to display the functional behavior of the TEGR angular momentum for the selected quasinormal-mode contributions, namely its radial accumulation, its oscillatory dependence for individual modes, and its dependence on the regularization parameter $\alpha$. In particular, for small values of $\alpha$, the corresponding profiles remain close to the $\alpha=0$ case.

The radial dependence of the perturbative gravitational angular momentum is shown in Figs.~1 and~2. 
Figure~1 displays $\delta J(R)$ for different even axial quasinormal modes characterized by the multipolar index $\ell$ and overtone number $n$, evaluated at fixed time. The oscillatory radial structure reflects the spatial profiles of the corresponding axial modes and the cumulative character of the radial integral defining $\delta J$. 
Figure~2 shows the radial behavior of $\delta J$ for the fundamental mode $(\ell=2,n=0)$ and different values of the regularization parameter $\alpha$. In the regime of small $\alpha$, the curves are nearly indistinguishable from the $\alpha=0$ case, indicating that the radial accumulation of gravitational angular momentum is weakly sensitive to the regularization of the central region.

The temporal evolution of $\delta J$ is presented in Figs.~3 and~4. 
In Fig.~3, $\delta J(t)$ is shown for different even axial quasinormal modes, exhibiting an oscillatory decay as expected for quasinormal frequencies. 
Figure~4 illustrates the time dependence of $\delta J$ for the fundamental mode and different values of $\alpha$. As in the radial case, small values of the regularization parameter lead to a time evolution that closely follows the $\alpha=0$ behavior, with differences restricted to mild modulations of the amplitude and damping timescale. This reinforces the conclusion that, within the present framework, regular and singular geometries may give rise to nearly degenerate signatures in the perturbative gravitational angular momentum.

\section{Conclusion}

In this work we derived a closed and remarkably simple expression for the
gravitational angular momentum carried by axial perturbations of the regular
Bardeen black hole within the teleparallel equivalent of general relativity.
Using the well-defined notions of energy and angular momentum provided by the
TEGR framework, the angular momentum was shown to be entirely determined by a
surface term and to obey a clear multipolar selection rule. In particular, only
even--$\ell$ axial modes contribute to the angular momentum, while odd--$\ell$
modes carry no net angular momentum. This result provides a new characterization of gravitational perturbations of
regular black holes and complements previous analyses based on quasinormal
frequencies. It also suggests that parity properties may play a nontrivial role
in the dynamics of polar perturbations, for which a complete quasinormal-mode
analysis is still lacking in the case of the Bardeen spacetime. Further
investigations along these lines may help clarify how conserved quantities are
distributed among different perturbative sectors of regular black holes.

It is worth emphasizing that the interpretation of ringdown observations in terms of a unique background geometry is intrinsically limited by degeneracies among different effective descriptions of black-hole spacetimes. Recent studies have shown that distinct geometrical models may give rise to nearly indistinguishable quasinormal-mode spectra within current observational uncertainties, even when multiple modes are considered~\cite{Shaikh2023mnras}. This limitation becomes particularly relevant in light of recent advances in black-hole spectroscopy based on gravitational-wave observations, which have demonstrated that the ringdown phase of binary black-hole mergers can be described with increasing precision in terms of a small number of quasinormal modes, as illustrated by the analysis of the event GW250114~\cite{GW250114PRL}. While such results are consistent with the Kerr description within the class of models tested, they do not establish the uniqueness of the underlying spacetime geometry. In particular, regular black-hole configurations that smoothly approach their singular counterparts may remain observationally indistinguishable within current experimental uncertainties, as illustrated by the nearly degenerate behavior of $\delta J(t)$ for small values of the regularization parameter in Fig.~4. In this context, the multipolar selection rule derived here for the gravitational angular momentum provides a concrete and complementary target for experimental tests of gravitational dynamics, going beyond standard spectroscopic consistency checks based solely on quasinormal frequencies.


\end{document}